\documentclass{elsart}

\usepackage{graphicx}
\usepackage{amssymb}

\begin{document}

\begin{frontmatter}

% Title, authors and addresses

% use the thanksref command within \title, \author or \address for footnotes;
% use the corauthref command within \author for corresponding author footnotes;
% use the ead command for the email address,
% and the form \ead[url] for the home page:
% \title{Title\thanksref{label1}}
% \thanks[label1]{}
% \author{Name\corauthref{cor1}\thanksref{label2}}
% \ead{email address}
% \ead[url]{home page}
% \thanks[label2]{}
% \corauth[cor1]{}
% \address{Address\thanksref{label3}}
% \thanks[label3]{}

\title{Signatures of Ultra-High Energy Cosmic Ray Composition
	from Propagation of Nuclei in Intergalactic Photon Fields}

% use optional labels to link authors explicitly to addresses:
% \author[label1,label2]{}
% \address[label1]{}
% \address[label2]{}

\author[label1]{T. Yamamoto}
\author[label2]{K. Mase}
\author[label4]{M. Takeda}
\author[label4]{N. Sakaki}
\author[label3]{M. Teshima}

\address[label1]{Center for Cosmological Physics, University of Chicago,
	 Chicago IL 60637 USA}
\address[label2]{Institute for Cosmic Ray Research, University of Tokyo,
 Chiba 277-8582, Japan}
\address[label3]{Max-Planck-Institute f$\ddot{u}$r Physik, F$\ddot{o}$ringer 
	Ring 6, 80805 M$\ddot{u}$nchen, Germany}
\address[label4]{RIKEN (The Institute of Physical and Chemical
 Research), Saitama 351-0198, Japan}

\begin{abstract}
% Text of abstract

We present a calculation of nuclei propagation with
energies above $10^{18}$ eV in the intergalactic photon field.
The calculation is based on a Monte Carlo approach for the
nucleus-photon interaction as well as the intergalactic magnetic field. 
We then assume that the Ultra-High Energy Cosmic
Rays (UHECR) are nuclei which are emitted from extra-galactic point sources. 
Four bumps are found in the energy spectrum of the UHECR which form clusters 
in the distribution of their arrival directions.
Based on this calculation,
the energy distribution of the clustered events is discussed.

\end{abstract}

\begin{keyword}
% keywords here, in the form: keyword \sep keyword
ultra high energy cosmic rays, propagation
% PACS codes here, in the form: \PACS code \sep code
\PACS 
\end{keyword}
\end{frontmatter}

% main text
\section{Introduction}
\label{sec:intor}

After a few decades of observation by several projects, 
the nature of the UHECR is gradually being revealed.
Based on the observational results, there is no
doubt of the existence of UHECR up to $10^{20}$ eV
\cite{agasa0,Flyseye,hires2002}.
This has led to consideration of a large number of production models. 
Most of the models assume that UHECRs are protons or photons.

Several studies of anisotropy have been done to reveal 
the origin of the UHECR
\cite{agasa1,uchiho,agasa3,farra1,farra2,tinya1}. 
Around $10^{18}$ eV, a large scale anisotropy which
is correlated to the Galactic center was reported by AGASA group \cite{agasa1}.
This result, if confirmed, supports an interpretation that 
cosmic rays below $10^{18}$ eV are dominated by a Galactic origin.
Above this energy, however, no significant large scale anisotropy 
has been detected by AGASA \cite{agasa3}.

On the other hand, AGASA reported clusters in the distribution of
arrival directions including 4 doublets and 2 triplets with energy
above $4\times 10^{19}$ eV\cite{uchiho,agasa3}. This results indicate
that the UHECR come from extra-galactic point sources. The cluster
events reported by AGASA are composed of particles with different
energies. A possible explanation would be neutral particles (like
neutrinos) propagating in straight lines from the source.  Another
possible explanation would be charged particles with large rigidity ,
i.e. with small charge like protons. 
%Protons at $10^{20}$ eV emitted by a
%source 100 Mpc away would be deflected a couple of degrees depending
%on the assumptions of the extragalactic magnetic field. Additionally,
%it was shown in \cite{Cronin} that AGASA clusters will not be
%distorted by the galactic magnetic field.  
If there is only the effect of the Galactic
magnetic field then protons of different energy can arrive at same
direction \cite{Cronin}.
However, these two
possibilities cannot explained the three bumps in the energy spectrum
of the cluster events between $10^{18.8}$ eV and $10^{19.8}$ eV
\cite{agasa4} reported by AGASA. This structure is not statistically
significant. However, it suggests to us the interesting possibility
that UHECR might be nuclei as it will be discussed in this paper.
In this scenario, cluster events would be formed by particles with 
different charge and energy but with the same rigidity.
The atomic number of the primary cosmic ray does not affect 
the energy determination of AGASA which measures the number of 
charged particles on the ground. However, the capability of AGASA to measure  
the composition of the primary cosmic rays is not very strong \cite{Kenjik}.
Nevertheless, the structure of the energy distribution of the cluster events 
may be used to measure of the composition indirectly 
as discussed in this paper.

If UHECR come from point sources and their rigidity is large enough or
their source close enough so the direction is not lost, a fraction of
the particles emitted by the source will arrive within a small angle
from the source position. These particles are what we will call in this
paper ``small deflection angle'' particles. Clusters will be formed
when the number of particles expected within this angle in our
experiment is larger than one.  The small deflection angle particles
should represent the cluster events observed by AGASA.

UHE protons should interact with Cosmic Microwave Background Radiation
(CMBR) and lose energy by photo-pion production. For this reason,
$10^{20}$ eV protons cannot propagate in intergalactic space beyond 50
Mpc, leading to a cut off in the spectrum below $10^{20}$ eV,
the so called GZK cut off.  On the other hand heavy nuclei will
interact during its propagation with the Intergalactic Infrared
Background Radiation(IIBR). Unfortunately the IIBR cannot be measured
directly so far because of the strong background of infrared photons
of Galactic origin.  The main component of the IIBR is thought to be
red-shifted stellar emission and reemission from interstellar
dust. The UHE nucleus should be disintegrated by the IIBR and lose
energy by the photo-disintegration interaction.  For this
reason, it was believed that the attenuation length of UHE nuclei is
shorter than that of protons in intergalactic space
\cite{steck2,steck4}.  However, recent calculations which are based on
empirical data show the IIBR density is much lower than
originally thought \cite{steck5}.  The lack of absorption in the
energy spectrum of very high energy gamma rays from AGNs strongly
supports this calculation. Using this result, the attenuation length
of the UHE nuclei in the intergalactic photon field was re-estimated
\cite{steck1} and found to be longer than that of a proton (a few 100
Mpc for $10^{20}$ eV Fe ).

We simulate the propagation of UHE nuclei which are emitted
from extra-galactic point sources based on a Monte Carlo approach.
Using this simulation, we found structures in the 
energy distribution of the UHE nuclei which are observed as cluster
events on the earth. 
In section \ref{sec:photons} we briefly describe the interaction 
of nuclei with background photons. In section \ref{sec:magnetic}
we describe the scattering of the nuclei by the intergalactic magnetic field.
In section \ref{sec:energy} we show the result of this simulation.
We conclude in section \ref{sec:discussion} and discuss the results
briefly.

\section{Interaction of Nuclei with background photons}
\label{sec:photons}

There are three processes which affect cosmic rays during
intergalactic propagation:
energy loss by interaction with cosmic background radiation;
deflection by magnetic fields; and energy loss by adiabatic expansion of
the universe. The effect of the adiabatic expansion is negligibly small 
if propagation distances are less than 1 Gpc.

\begin{figure}
\includegraphics[width=1.0\textwidth]{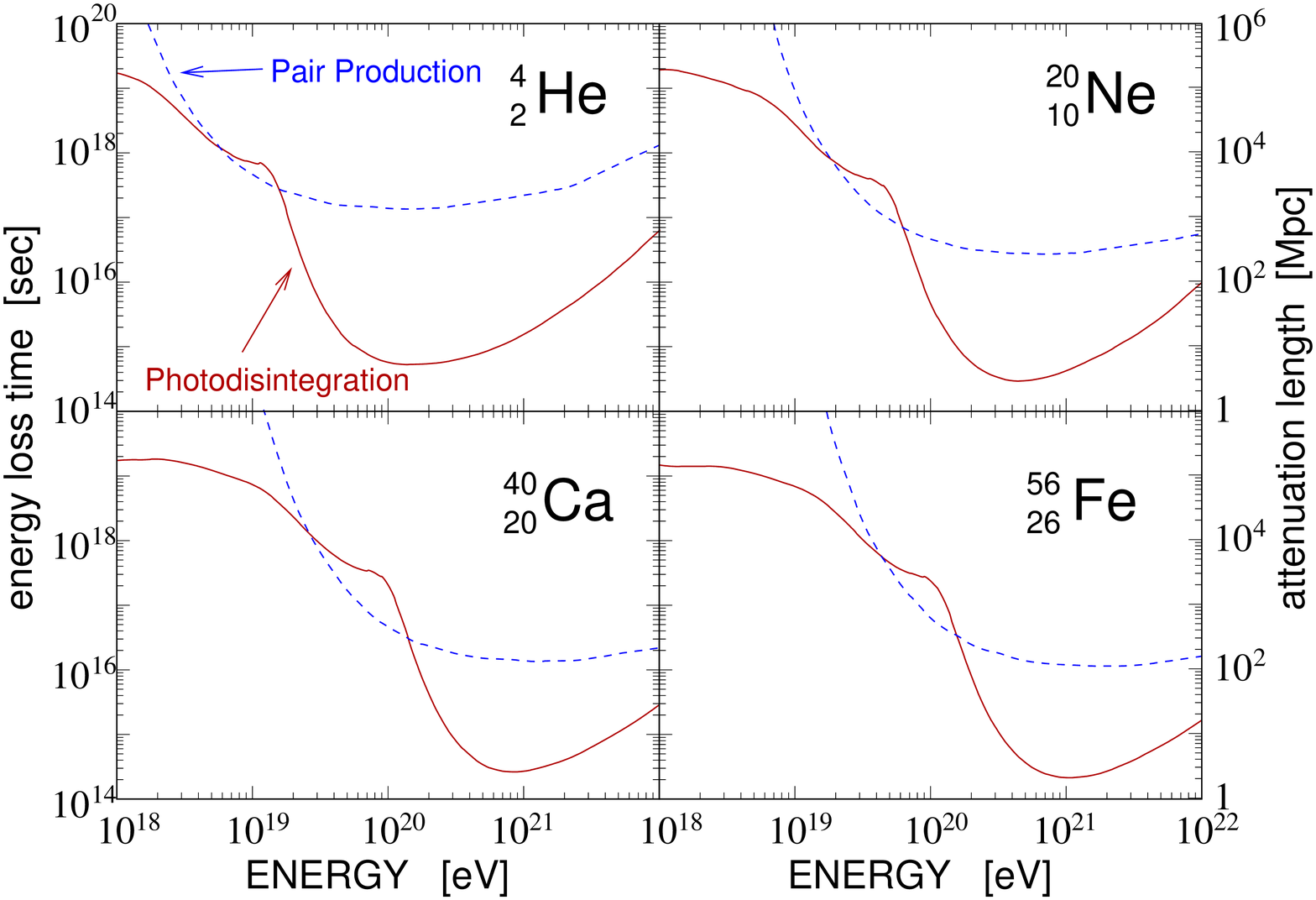}
\caption{Energy loss time of different mass nuclei as a function of energy. 
Solid line is that of single-nucleon emission by photo-disintegration 
and dashed line is pair production \cite{steck1}.}
\label{fig:f2}
\end{figure}

%Most of the energy loss by the cosmic background radiation is due to the CMBR.
%(Stecker \& Salamon 1999\cite{steck1}).
There are four interaction processes of cosmic rays with intergalactic
 background radiation.  The first process, Compton interaction, is
 negligibly small for our purpose.  The second process is
 pair-production of e$^+$e$^-$.  This interaction can occur if the
 energy of the background photon is greater than 1 MeV in the rest
 system of the nucleus. If the Lorentz factor of the nucleus is larger
 than $10^{9}$, it will efficiently interact with the CMBR 
(Figure \ref{fig:f2}).  In this
 case, the energy loss rate ($dE/dt$) is proportional to the square of
 the charge of the nucleus.  The third process is photo production
 which mainly affects protons with energy greater than $10^{20}$ eV.
% as photo-pion production which causes the GZK cut off. 
The interaction of heavier nuclei with the CMBR by this process is
negligible for the energy range considered in this paper.

The fourth process is photo-disintegration. 
%The calculation of this
%photo-nucleus interaction was originally reported in the reference of
%\cite{steck1}.  
The nucleus interacts with the background photon and
disintegrates to a lighter nucleus.  In this paper, we calculate this
process by a Monte-Carlo method according to \cite{steck4,steck1}.  These
authors parametrized the total cross section $\sigma(\epsilon)$ as a
function of photon energy in the rest frame of the nucleus.  Then the
probability of photo-disintegration per unit length $R$ is calculated
by following equation:
\begin{equation}
R=\int_0^{\infty} n(\epsilon) \left[ \frac{\int_0^{2\gamma\epsilon}
	\sigma(\epsilon^\prime) \frac{\epsilon^\prime}{\epsilon}
	d \epsilon^\prime}{2 \gamma \epsilon} \right]
	d \epsilon
\end{equation}
where $\epsilon$ and $\epsilon^{\prime}$ are the energies of the photon 
in the lab frame and in the nucleus rest frame respectively,
$n(\epsilon)$ is the differential number density of photons including the 
CMBR and the IIBR \cite{steck1},  
and $\gamma=\epsilon^{\prime}/ \epsilon$ is the Lorentz factor of the nucleus.
The term inside the bracket corresponds to the angle-averaged cross-section 
for a photon of energy $\epsilon$.
Figure \ref{fig:f2} shows characteristic time $\tau$ for the energy loss 
due to photo-disintegration 
for a single-nucleon emission process as a function of energy.
$\tau$ is defined by following equation:
\begin{equation}
\frac{1}{\tau}=\frac{1}{E}\frac{dE}{dt}
\end{equation}

The effect of the CMBR is maximized when the Lorentz factor $\gamma$ is 
around $10^{10}$. In case of $Fe$, this value corresponds to
an energy of $10^{21}$ eV, yielding a value of 
$\tau = 2\times 10^{14}$ s (2 Mpc). 
Because the IIBR density is much lower than the CMBR density,
$\tau$ increases rapidly at lower energy.
$\tau$ of $Fe$ with $10^{20}$ eV is about $2.5\times10^{17}$s (2.5Gpc). 
In general, one or a few nucleons and a single lighter nucleus are
emitted by this interaction. The Lorentz factor is conserved at each 
photo-disintegration interaction, though it is reduced by the pair-production. 
Finally, UHE particles will pile up at a Lorentz factor around
$10^{9}$. When $^{9}{Be}$ is disintegrated, one proton and
two $He$ are emitted. Therefore no nucleus is created with a mass number 
between 5 and 8.

The interaction probability is calculated for 
every possible photo-disintegration process involving
the emission of one or more nucleons for all nuclei lighter than $Fe$.
Based on these probabilities, the propagation of the nucleus is
simulated by the Monte-Carlo method. The effects of pair production, 
the adiabatic expansion, and photo-pion production for protons are taken
into account analytically in this simulation. 

\begin{figure}
\includegraphics[width=1.0\textwidth]{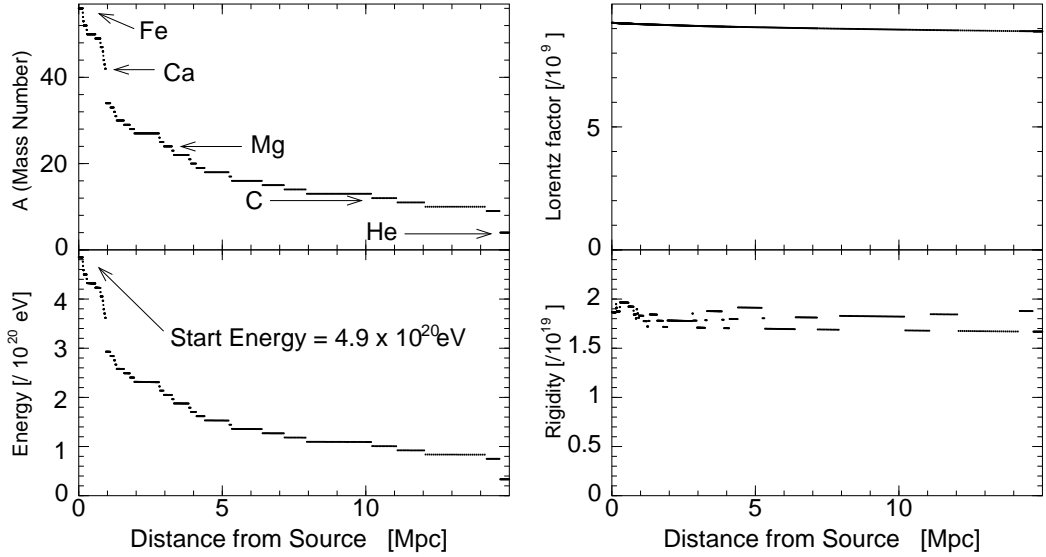}
\caption{Life of an $Fe$ nucleus. An $Fe$ is emitted with an energy of 
$4.9\times 10^{20}$ eV. The $Fe$ is disintegrated during its propagation
and turn into lighter nucleus with the emission of nucleons.
These figures show the properties of the surviving nucleus as a function of
propagation distance from the source. The upper and lower left panels 
show the variation of the mass number and the energy of the surviving nucleus
respectively.
These parameters change in a similar way. The upper and lower right panels show
the variation of the Lorentz factor and the rigidity respectively.
In comparison with the energy, the Lorentz factor changes smoothly because 
the photo-disintegration process does not change the Lorentz factor. Only
the pair-production process affects this parameter. Therefore variations of
the Lorentz factor and the rigidity are smaller than that of the energy.
The fluctuation of the rigidity in lower right panel is caused by
the variation of the ratio of mass number to charge(A/Z).
}
\label{fig:life_of_fe}
\end{figure}

Figure \ref{fig:life_of_fe} shows the life of an $Fe$ nucleus
whose initial energy is $4.9\times 10^{20}$ eV. The $Fe$ is disintegrated
during its propagation and turns into a lighter nucleus with the emission 
of nucleons. When the nucleus is $^{42}Ca$, four protons and four 
neutrons are emitted. Then a $Sulfur$ is created. In comparison with 
energy, the Lorentz factor changes smoothly because photo-disintegration 
process does not change the Lorentz factor. This process makes a lighter 
nucleus and nucleons with same Lorentz factor.

\section{Scattering by the intergalactic magnetic field}
\label{sec:magnetic}

The effect of magnetic fields is described by the rigidity 
defined as the ratio of energy to charge ($=E/Z$). Particles
with small rigidity are deflected by the magnetic field and cannot be observed 
as a cluster. The deflection of magnetic field increases
the propagation time, and therefore particles emitted from distant 
sources may not reach to the observer.
We calculate the effect of the intergalactic magnetic
field based on a Monte Carlo method. 
%In this section, the effect of
%magnetic field is described.

%1, Kolmogorov Extra-galactic magnetic field.\\

\begin{figure}
\includegraphics[width=1.0\textwidth]{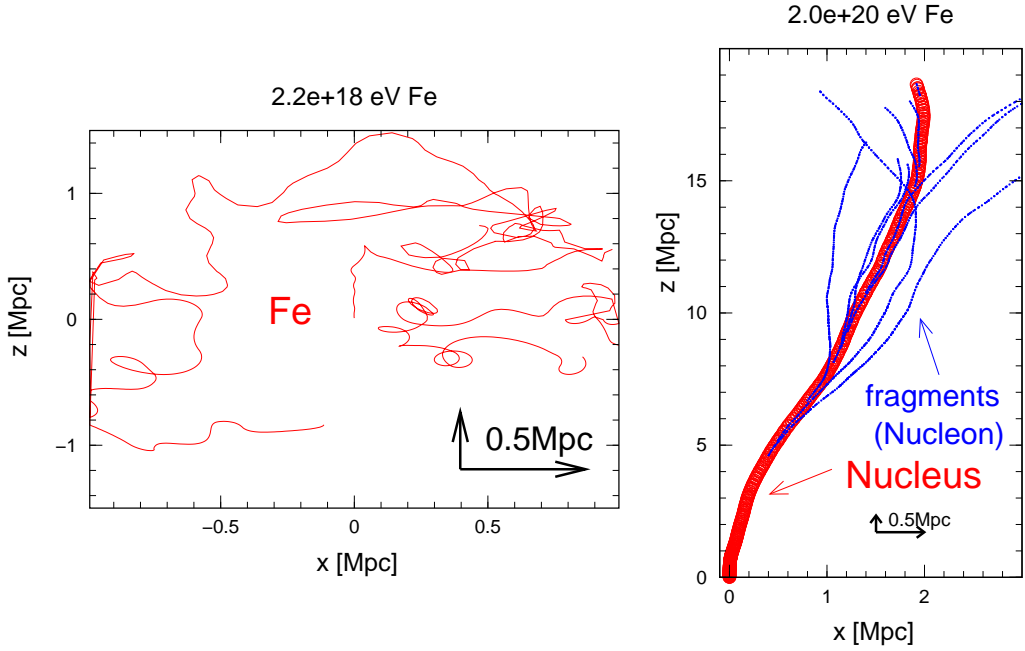}
\caption{Examples of trajectory of UHE $Fe$ in the
 intergalactic magnetic field. A $Fe$ emitted in the $z$ direction 
with an energy of $2\times 10^{18}$ eV from a source located on
the origin in left panel. A projection of the trajectory on x-z plane is shown.
At this energy, the attenuation length is sufficiently large. 
Therefore the energy of the particle is almost conserved and the particle
is scattered by the magnetic field significantly. The $Fe$ cannot propagate
a distance of more than 1 Mpc from the source. In the right panel, a $Fe$
with energy of $2\times 10^{20}$ eV emitted in the same direction as the
left panel.
The $Fe$ and secondary nuclei are indicated by the wide line and
secondary nucleons (proton and neutron) are indicated by the narrow lines. 
The $Fe$ interact with photons rapidly and is disintegrated to
lighter nuclei with emission of protons and neutrons. All of the particles are
scattered by the magnetic field but deflection angle is comparatively
small. Arrows in the panels indicate 0.5 Mpc distance.}
\label{fig:trace2}
\end{figure}

To simulate the scattering by the intergalactic magnetic field, we
assume a Kolmogorov spectrum for the random magnetic field according to
the reference of \cite{stanev}. In this reference, 
authors divide space in a lattice of 250 kpc cubes. The lattice is filled
with a random magnetic field which follows the Kolmogorov spectrum with three
logarithmic scales. Three field vectors of random orientation are sampled
at scales $l$ = 1000, 500, and 250 kpc with amplitudes proportional to
$l^{1/3}$. The final magnetic field in each 250 kpc cube is 
vectorial sum of these three vectors. The average magnitude of the magnetic 
field is assumed to be 1 nG. Particles propagate in spiral
trajectories until they leave the lattice. 
Figure \ref{fig:trace2} shows
examples of the trajectories. $Fe$ with energy of $10^{18}$ eV does not 
rapidly lose energy by the interaction with photons, and is trapped by the
magnetic field inside 1 Mpc cubes. In case of $2\times10^{20}$ eV, $Fe$ is
disintegrated by the photons rapidly and the products propagates close
to a straight line in the initial direction of the $Fe$.

\section{Energy Distribution}
\label{sec:energy}

In our simulation, the source energy spectrum $dN(E)/dE$ is assumed
to have a power law dependence $E^{-2}$ and have a cut-off at the energy
of $Z \times 2\times 10^{19}$ eV, where $Z$ is charge of each primary nucleus.
Below 300 MeV/nucleon, the cosmic nuclear composition can be divided to 
4 types: $He$, $CNO$, $Ne-Si$, and $Fe$. Nuclear abundance 
between $He$ and $C$ is relatively low as 
well as between $Si$ and $Fe$. We treat the composition as 4 components, 
$_{4}^{2}He$, $_{14}^{7}N$, $_{24}^{12}Mg$, and $_{56}^{26}Fe$. 
The fraction of these nuclear abundances in the UHE region are unknown. 
Therefore we assume equal fractions relative to $He$ at the source.
20000 particles are created for each nucleus.

\begin{figure}
\includegraphics[width=1.0\textwidth]{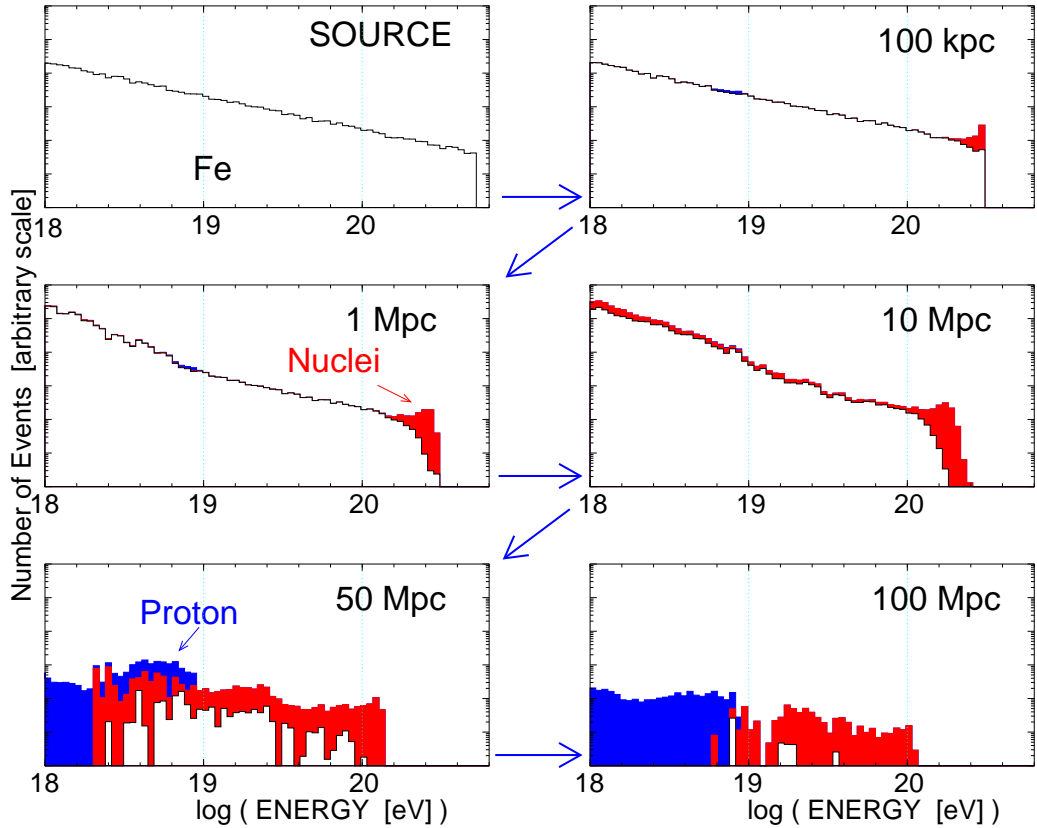}
\caption{
Variation of energy distribution of particles that began as $Fe$
after propagation over various distances from a single source.
The distances from the source are indicated in each panel. 
The number of particles (arbitrary scale) are
shown as a function of energy. Primary nuclei ($Fe$), 
secondary lighter nuclei, and protons are
indicated by white, light shade and dark shade in the histograms respectively.
The differential source spectrum is assumed
to be a power law, $\propto E^{-2}$, with energy
cut-off at $Z \times 2 \times 10^{19}$ eV (upper left panel). 
At 100 Mpc from the source, most of the $Fe$ disappear because of
their low rigidity.
}
\label{fig:180_d_fe}
\end{figure}

Figure \ref{fig:180_d_fe} shows the variation of the energy distribution 
after propagation from a single source initiated by $Fe$ through 
the photon and magnetic field.
Nuclei with Lorentz factor above $10^{9}$ interact with the CMBR  
rapidly. After 1 Mpc propagation a small pile-up is induced below the 
corresponding energy and the maximum energy of the particle is reduced.
Since we assume a cut-off energy in the source spectrum, protons 
are emitted with energies below $Z/A \times 2 \times 10^{19}(\simeq 10^{19})$ 
eV where $A$ is mass number of the primary nucleus.
The number of low energy nuclei ($<10^{19}$eV) increases up to 10
Mpc from the source since these nuclei are trapped by the magnetic
field. These low energy nuclei cannot propagate farther than few 10
Mpc. Therefore, the energy distribution becomes steeper for distances
 below 10 Mpc and flatter for larger distances.
No particles remain after propagation of a few 100 Mpc.

\begin{figure}
\includegraphics[width=1.0\textwidth]{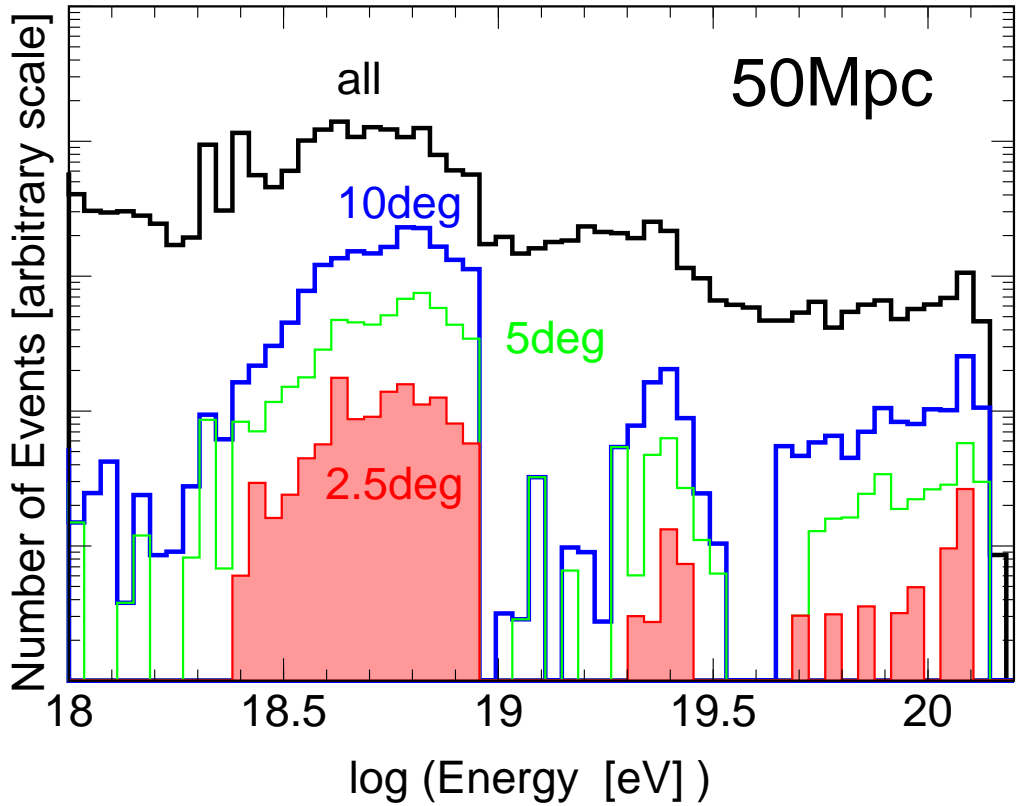}
\caption{
Energy distribution of the particles at the 50 Mpc distance from a source
as a function of deflection angle by the magnetic field.
The assumed source spectrum is same as in the previous figure.
The number of particles with the deflection angle smaller than 2.5, 5,
10, 180(same as previous figure) degrees are shown.
}
\label{fig:psrc_50mpc}
\end{figure}

Figure \ref{fig:psrc_50mpc} is same result as lower left panel of
Figure \ref{fig:180_d_fe}. The energy distributions at a distance of
50 Mpc from the source are shown for different 
deflection angle (angle between the arrival direction and the source position).
Particles with the deflection angle smaller than 2.5, 5, 10, 
180 degrees are counted (180 degrees corresponds to the entire sample of 
particles included in Figure \ref{fig:180_d_fe}).
Three bumps appear clearly in the distribution of 
particles with a small deflection angle.
The bump below $10^{19}$ eV is composed of protons which
are emitted from higher energy nuclei in the vicinity of the observer
who is located at 50 Mpc distance from the source. 
The bump around $10^{19.4}$ eV is composed of $He$, and the bump
above $10^{19.6}$ eV is composed of nuclei heavier than $Be$.

These bumps can be explained as follows:
The rigidity is proportional to $Z^{-1}$ and $E$, and $Z$ is approximately
proportional to its mass number $A$ while the Lorentz factor is proportional
to $A^{-1}$ and $E$. This means that the Lorentz factor of a particle is 
approximately proportional
to the rigidity. Therefore, if the Lorentz factor of the nucleus is large,
the energy is reduced by the photo-disintegration. If the Lorentz
factor is small, the particle is deflected by the magnetic field due to
its small rigidity. However particles with Lorentz factor of about
$10^9$ remain after applying the deflection angle cut.
As a result, the bumps appear in the energy distribution of the small 
deflection angle particles at positions dependent on the mass numbers.
The combination of the photodisintegration and the magnetic 
field acts as a ``filter'' for $\gamma=10^{9}$ particles.
The energy distribution of the small deflection angle particles
shows clearly the spectral features created by interactions with 
the intergalactic photon background and magnetic fields during their 
propagation.

%%%%%%%%%%% integration %%%%%%%%%%%%%%%%%%%%%
If the sources are distributed uniformly in the Universe, the expected
energy distribution measured at the earth can be estimated by adding
up linearly the energy distributions corresponding to different
distances from the source. In Figure \ref{fig:180_d_fe} the number of
particles around $10^{20}$ eV coming from sources at 100 Mpc is 1.5
orders of magnitude smaller than from sources at 100 kpc.  When adding
linearly the different distances, the number of particles from sources
at 100 Mpc is enhanced 3 orders of magnitude compared to sources at
100 kpc. Therefore, the contribution to the energy spectrum at earth
is increasing with the distance to the source till a couple of hundred
Mpc. On the other hand, most of the low energy particles emitted by
distant sources will be trapped by the magnetic field and they will
not reach the earth. However, the low energy secondaries resulting
from the photodisintegration of high energy heavy nuclei close to
earth will regenerate the low energy spectrum coming from distant
sources.  Therefore, even at low energies we will still be dominated by 
the contribution of distant sources.
The spectrum of small deflection angle particles will follow the same
behavior, i.e. the contribution of sources at large distances to 
the spectrum exceed largely the one from small distant sources, 
even if the fraction of small deflected particles for closer 
sources is slightly larger.

It should be noted that the relation between cluster events and small
 deflection particles is qualitatively clear but difficult to estimate
 quantitatively. Current work is in progress for a better comparison with
 experimental data.
%%%%%%%%%%%%%%%%%%%%%%%%%%%%%%%%%%%%%%%%%%%%%

\begin{figure}
\includegraphics[width=1.0\textwidth]{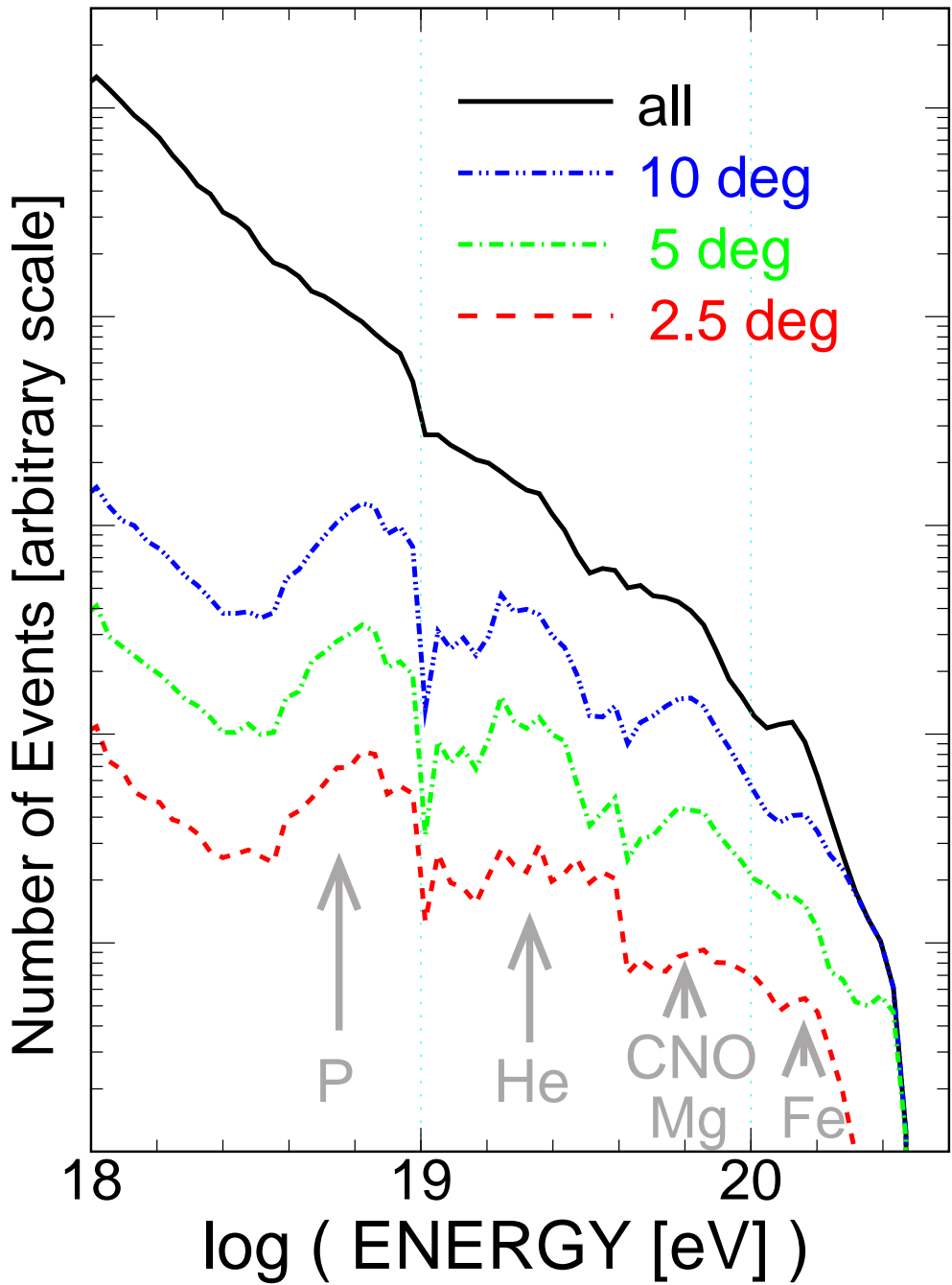}
\caption{The expected energy distribution of particles after
propagation under the
 assumption that point sources are distributed uniformly in the
 universe. The solid line shows the energy distribution with no
 deflection angle cut. The other (dashed dots) lines show the energy 
distributions of the particle with deflection angles smaller than 10, 5, 
and 2.5 degrees. 
 Particles with lower rigidity are scattered by the magnetic field.
 Larger rigidity particle are disintegrated by the photon field
 due to their large Lorentz factor. }
\label{fig:spect_all}
\end{figure}

Figure \ref{fig:spect_all} shows the result of the expected energy distribution
under the assumption that the sources are distributed uniformly in the
Universe. 
The solid line shows the expected energy distribution of all  the particles
that reach the earth. The expected energy spectrum is steeper than
the source spectrum. Above $10^{20.2}$ eV, the energy spectrum becomes even
steeper and the maximum energy is about $10^{20.4}$ eV.
The exact details of the upper end of the spectrum
will depend on the source spectrum cut off.

The dashed lines in Figure \ref{fig:spect_all} show the expected energy 
distribution of the small deflection angle particles. This spectrum is
much flatter than the source spectrum. 
Four bumps appear in this figure. The bump just below $10^{19}$ eV 
corresponds to protons and strongly depends on the cut off
energy at the source. The bump around $10^{19.3}$ eV is composed of
$Deuteron$ and $He$. Every nucleus with a Lorentz factor around
$10^9$ contributes to the $He$ bump after propagation. The bump
between $10^{19.6}$ eV and $10^{20}$ eV correspond to nuclei with mass
between $Be$ and $Mg$. The bump above $10^{20.1}$ eV is composed of $Fe$ and
other nuclei of similar atomic number.

This total spectral structure bears a striking resemblance to the
result of AGASA \cite{agasa0,agasa2,agasa4}.

\section{Results and Discussion}
\label{sec:discussion}

We have simulated propagation of UHE nuclei which are
emitted from extra-galactic point sources using a Monte Carlo 
approach. Using this simulation, structures in the energy distribution 
of the small deflection angle particles have been predicted.
In this simulation, we assume that UHECR are nuclei which are
emitted from extra-galactic point sources. The nuclei interact with
cosmic background radiation and lose energy during their
propagation. Based on recent estimations of the IIBR, it has been shown that 
UHE nucleus can propagate longer distances than a proton with same
energy. It is also expected that heavier nuclei are accelerated more
efficiently to high energy than lighter nuclei or protons.

We have considered the effect of the intergalactic magnetic fields. We
use a Kolmogorov spectrum for random magnetic fields that is 1 nG in
average intensity with a 1 Mpc correlation length. 
The energy distribution of the
particles with deflection angles smaller than 10, 5, and 2.5 degrees are
investigated. Consequently, for this model no cosmic rays can be observed 
after propagation of a few hundred Mpc. Lower rigidity particles are deflected
by the magnetic field relatively rapidly. Particles with larger Lorentz factor 
lose energy by photodisintegration. Since the rigidity and the Lorentz factor 
are proportional to the particle energy and inversely proportional to
the mass number, particles which have common a Lorentz factor, 
or similar rigidity, remain after the deflection angle cut. 
The $Be-Mg$ bump appears at an energy more than 4 times larger than the 
$He$ bump, and the $Fe$ bump appear around 6 times
larger energy. The rigidity gives a lower bound and energy loss gives an upper
bound on the bumps. The energy distribution of the clustered events observed by
AGASA can be explained by this model.

In the simulation, we assume a specific intergalactic magnetic field
distribution, cut-off energy and source composition of primary nuclei.
The Galactic magnetic field is ignored in this simulation.
The events with certain deflection angle are selected and shows 
the effects due to the intergalactic photons and magnetic fields. 
This effect depends on the strength and correlation of the
intergalactic magnetic field. This dependence gives
an ambiguity to the analysis since the deflection angle cut
is adjustable. Therefore, this angle should be treated carefully.

This result suggests that if we detect few hundred events above $10^{20}$ 
eV in future experiments, we should be able to extract the structure of 
the bumps in the energy distribution of cluster events. The Pierre Auger
Observatory is quite sufficient for this purpose and 
may confirm the signature of nuclei in the spectrum 
in the near future.

\section*{Acknowledgments}

We thank Angela Olinto, Alan Watson, James W. Cronin, Maximo Ave,
Aaron Chou, and Medina Tanco for useful discussions.
This research was supported by grant NSF PHY-0114422 to the Center for
Cosmological Physics, University of Chicago, and grant NSF PHY0103717.

% The Appendices part is started with the command \appendix;
% appendix sections are then done as normal sections
% \appendix

% \section{}
% \label{}

\newpage

%%%%%%%%%%%%%%%%% figure %%%%%%%%%%%%%%

%\begin{figure}
%\includegraphics[width=1.0\textwidth]{angle_2.eps}
%\caption{
%The variation of deflection angle as a function of source distance.
%Particles with energy of above and below $10^{19.5}$ eV are
%compared. The dotted lines indicate a uniform distribution for reference.
%The source distances are indicated in each panel.
%At 50 Mpc most of the lower energy particles are emitted from higher
%energy particles in the vicinity of the observer.
%}
%\label{fig:angle_2}
%\end{figure}

\end{document}